\documentclass[conference]{IEEEtran}
\usepackage{cite}
\usepackage{amsmath,amssymb,amsfonts}
\usepackage{algorithmic}
\usepackage{graphicx}
\usepackage{textcomp}
\usepackage{xcolor}
\usepackage{hyperref}
\hypersetup{
    colorlinks=true,
    linkcolor=blue,
    citecolor=blue,
    urlcolor=blue,
}
\usepackage{float}
\usepackage{caption}
\usepackage{booktabs}
\usepackage{array}

\usepackage{fancyhdr}
\fancypagestyle{firstpage}{
    \fancyhf{} % Clear default headers and footers
    \fancyfoot[L]{Confidential — Internal Use Only}
    \fancyfoot[R]{Page \thepage}
}

\usepackage{float}
\let\origfigure\figure
\let\endorigfigure\endfigure
\renewenvironment{figure}[1][H]{\origfigure[H]}{\endorigfigure}

\makeatletter
\def\fps@figure{H}
\makeatother

\begin{document}

\title{Optimization of a Solar-Grid Water Supply System with Hydraulic Energy Storage: A Case Study}

\author{\IEEEauthorblockN{Mariamawit Gashahun Arega}
\IEEEauthorblockA{
\textit{College of Engineering}\\ \textit{Carnegie Mellon University}\\
Kigali, Rwanda\\
marega@andrew.cmu.edu}
\and
\IEEEauthorblockN{Jesse Thornburg}
\IEEEauthorblockA{
\textit{College Of Engineering, Carnegie Mellon University, PA, USA}\\
Research Fellow, Kigali Collaborative Research Centre, Kigali, Rwanda \\
Faculty Affiliate, Scott Institute for Energy Innovation at CMU, PA, USA  \\
jthornbu@andrew.cmu.edu
}
}
% Source - https://tex.stackexchange.com/a/618225
% Posted by Engr. Nnoli K. Pal, modified by community. See post 'Timeline' for change history
% Retrieved 2026-11-22, License - CC BY-SA 4.0

\IEEEoverridecommandlockouts
\IEEEpubid{\makebox[\columnwidth]{*979-8-3195-1706-7/26/\$31.00~\copyright2026 IEEE \hfill}
\hspace{\columnsep}\makebox[\columnwidth]{ }}

\maketitle

\begin{abstract}
Given the high cost of grid-scale batteries, other energy storage approaches are needed for developing and low-income regions. Elevated volumes of water are often more economically feasible energy storage options than batteries in these contexts. This paper presents a dual-framework modeling approach for one such energy storage setup: a solar-grid hybrid water supply system in the arid Afar region of Ethiopia. By utilizing an elevated 1,900 m³ water reservoir as a "hydraulic battery," the system replaces traditional electrochemical storage, significantly reducing capital expenditures. Using Python for high-resolution technical simulation and Xendee for economic optimization, the study shows a 55.1\% solar fraction and a 10\% Internal Rate of Return (IRR). The results confirm that hybrid systems with hydraulic storage offer a financially viable and reliable solution for water access in energy-constrained regions.
\end{abstract}
\vspace{-6pt} % Adjust the value (e.g., -1em, -15pt) as needed

\section{Introduction}

Sub-Saharan Africa faces a significant energy-water gap, with nearly half of the population lacking modern energy. In the Afar region only 4.1 per cent of people have access to improved sanitation facilities, while only 11\% use improved water sources \cite{unicef2022}. A situation exacerbated by frequent droughts and a highly unreliable national grid. Despite these challenges, the region possesses abundant solar resources, with an annual direct normal irradiation of approximately 2,045 kWh/m²\cite{worldbank2019}. As other recent research shows, pairing onsite solar PV with an erratic national grid can improve stability and reliability for local loads in East Africa \cite{ahereza}. Solar PV is an increasingly affordable generation technology that is ideal for much of sub-Saharan Africa since it can be sized and installed without costly and lengthy feasibility studies, even with automated sizing using satellite images and computer vision \cite{dossantos}. A recent study showed prefeasibility analyses for small-scale solar PV installation in East Africa can be completed with large language model (LLM) tools available free to the public online \cite{moses}.

This study presents the design and validation of a solar-grid hybrid water supply system. To optimally size it, the work implements a dual-framework approach using a meteorological model and Xendee optimization to verify the technical and financial viability of the microgrid. The design's technical robustness is also shown, with 100\% water delivery reliability and a solar fraction over 55\%. This method is demonstrated in a case study of solar PV paired with established water reservoirs and tariff mechanisms in Afar, Ethiopia. The specific contributions are threefold: (i) a physics-based Python model that converts climate-sensitive water demand into an hourly electrical load profile; (ii) the explicit treatment of an existing elevated reservoir as an equivalent battery energy storage system (BESS), or "hydraulic battery," whose parameters are derived from gravitational-storage physics rather than assumed; and (iii) integration of this physical model with the Xendee techno-economic platform to jointly evaluate cost, solar fraction, and resilience under grid-outage conditions.

\begin{figure}[H]
\centering
\includegraphics[width=0.45\textwidth]{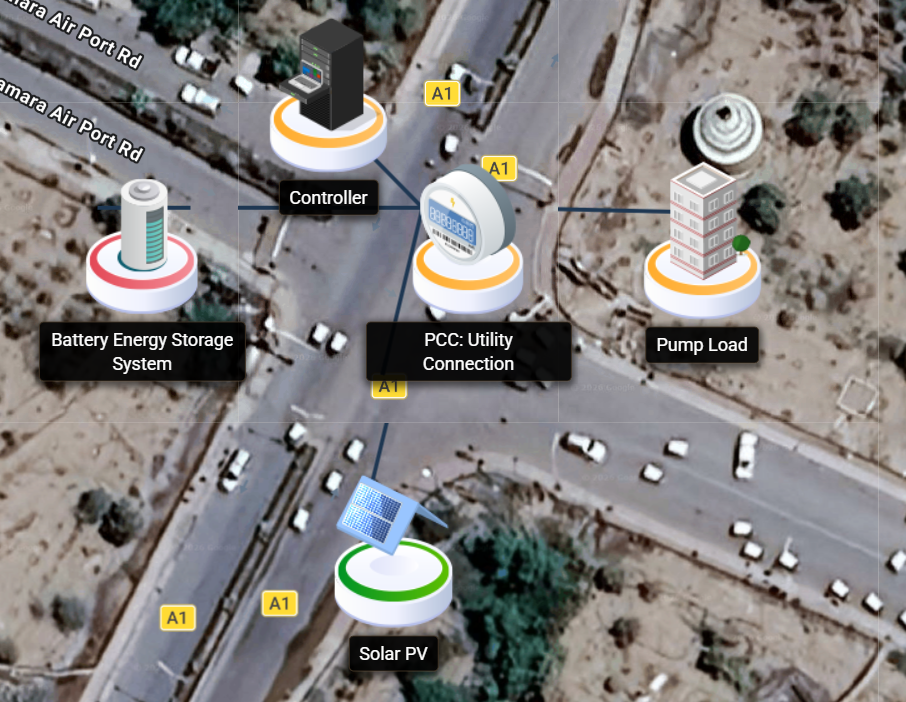}
\caption{System layout for Afar, Ethiopia, modeled in Xendee.}
\label{fig:xendee}
\vspace{-6pt} % Directly pulls the following text upward
\end{figure}

\section{Background}

Globally, 1.18 billion people live in energy poverty (unable to reliably use electricity), with around half residing in Sub-Saharan Africa \cite{akinyele2016}. Electricity access is a key contributor to any country's economic development, poverty reduction, and cleaner environment \cite{erbato2012}. Even relatively electrified nations such as South Africa still face a 20\% household electrification gap \cite{patel2018}. Renewable resources are abundant across the region and offer lower maintenance and operational costs than nuclear or fossil-based systems, yet rural electrification increasingly relies on decentralized microgrids that face persistent technical, economic, and policy challenges.

Prior studies illustrate the range of these challenges. Babayomi and Okharedia introduce the CETEP framework, emphasizing community ownership, financial viability, and policy uncertainty as determinants of long-term sustainability, but do not propose engineering-design adjustments responding to these social constraints \cite{babayomi2019}. Economic viability is also shaped by tariff design: Raji et al. show that time-of-use, block, and subscription tariffs directly influence consumption patterns, often with unintended consequences, yet such tariff studies are typically modeled independently of technical optimization \cite{raji2022}. At the integration level, Khan et al. identify gaps in Ethiopia's grid code—voltage ride-through, frequency response, and reactive power—that limit renewable penetration relative to IEEE 1547 standards \cite{khan2022}. Distribution-level decisions matter as well; Ehnberg et al. show that cable sizing in Ethiopian and Rwandan PV microgrids is shaped by local tariffs and carbon intensity, with oversizing beneficial in Rwanda but discouraged by Ethiopia's subsidized prices \cite{ehnberg2024}.

On system sizing, Abebe et al. used HOMER to optimize a hybrid PV–wind–battery microgrid in Gedro Kebele, Ethiopia, finding solar PV contributed 94\% of generation with wind only 6\%, underscoring the dominance of solar resources in Ethiopian contexts \cite{ashetehe2019}. Other studies integrate micro-hydro \cite{mekonnen2021} or irrigation infrastructure \cite{emmanouil2021}, while Beyene et al. explicitly model outage frequency and reliability constraints in industrial parks \cite{bekele2023}. Across this literature, however, most studies isolate technical feasibility from socio-economic realities, or policy from engineering design; few integrate technical reliability, economic sustainability, and community familiarity into a single framework. Our study addresses this gap with a solar-grid hybrid water supply system that reuses existing reservoirs, pumps, and tariff structures while introducing solar PV as a low-maintenance innovation, situating the design at the intersection of technical feasibility and social acceptance.

\section{METHODOLOGY}

\subsection{Integrated Modeling Framework}

This study adopts a hybrid simulation–optimization framework combining a Python-based physical system model with Xendee's time-series optimization platform to evaluate system performance, cost, and resilience. Python has been used in other studies for modeling optimization in East African energy projects, e.g., dispatch of hybrid solar-grid capacity for an installation in Rwanda \cite{ogbogu}. For the Afar study, a Python module generates an hourly electrical load profile for the water pumping system, used as a primary input into Xendee. The workflow has three sequential stages: (1) physical system simulation (Python), (2) time-series optimization (Xendee), and (3) economic and resilience assessment. This ensures that operational constraints derived from physics-based modeling are preserved during economic optimization.

\subsubsection{Rationale for Selecting Xendee}

Xendee was selected over general-purpose tools such as HOMER and custom solvers for multiple reasons. First, it performs mixed-integer linear programming on user-supplied hourly time series, allowing us to feed in the externally computed, physics-based load and PV profiles from our Python model rather than relying on internal simplified estimators. Second, it natively models a generic battery storage asset with configurable capacity, charge/discharge efficiency, and state-of-charge limits, which maps directly onto our "hydraulic battery" abstraction. Third, it provides built-in distribution-level economic and resilience reporting (NPV, IRR, tariff sensitivity, islanding), enabling joint techno-economic and reliability evaluation in one environment. HOMER offers less flexibility in importing fully exogenous load profiles, while a custom solver would have duplicated capabilities Xendee already validates. More generally, for simultaneous economic and technical modeling in a single software (without the need to model transients, etc. in transmission or distribution infrastructure) meant Xendee was ideal for this study rather than other power system simulators like Adaptricity, PowerFactory, GridLab-D, or PLASMiS \cite{9844342,powerafrica2021,thesis2018}.

\subsection{Python-Based Water–Energy System Model}

\subsubsection{Time-Varying Water Demand Modeling}

Water demand is modeled as a dynamic, time-dependent variable incorporating seasonal variation and temperature sensitivity, based on regional empirical data from Timotewos et al. for the Kola climatic zone (representative of Afar) \cite{timotewos2022}. Annual demand is distributed across three climatic seasons using normalized monthly multipliers: Belg (Feb–May), 34.5\% of annual demand, multiplier 1.035; Bega (Jun–Sep), 33.3\%, multiplier 0.999; and Kiremt (Oct–Jan), 32.2\%, multiplier 0.966. Monthly demand is then:

\begin{equation}
D_{month} = D_{avg} \times M_{season}
\end{equation}
where $D_{month}$ is monthly water demand (m³), $D_{avg}$ is average monthly demand (m³), and $M_{season}$ is the dimensionless seasonal multiplier. Demand is further adjusted for temperature using an elasticity coefficient:

\begin{equation}
D(t) = D_{base}(t) \times \left(1 + \alpha \cdot (T(t) - T_{ref})\right)
\end{equation}
where $D(t)$ is temperature-adjusted demand (m³) at time $t$, $D_{base}(t)$ is base demand (m³), $\alpha = 0.0528$ is the temperature elasticity coefficient (5.28\% per °C), $T(t)$ is ambient temperature (°C), and $T_{ref}$ is the reference temperature (°C). The final hourly demand is:

\begin{equation}
D_{hour}(t) = D_{daily} \times f_{diurnal}(t) \times M_{season} \times \left(1 + \alpha \Delta T(t)\right)
\end{equation}
where $D_{hour}(t)$ is hourly demand (m³), $D_{daily}$ is total daily demand (m³), $f_{diurnal}(t)$ is the normalized hourly usage fraction capturing morning and evening peaks (dimensionless), and $\Delta T(t) = T(t) - T_{ref}$ (°C).

\subsubsection{Load Generation and Pumping Demand}

The Python module converts time-varying demand into electrical load via hydraulic energy relationships. Pump power is:
\begin{equation}
    P = \frac{\rho g Q H}{\eta}
\end{equation}
where $P$ is pump electrical power (W), $\rho = 1000$ kg/m³ is water density, $g = 9.81$ m/s² is gravitational acceleration, $Q$ is flow rate (m³/s), $H$ is total dynamic head (m), and $\eta$ is the dimensionless pump efficiency \cite{elwakil1984}. Hourly energy consumption is:

\begin{equation}
E_{hour} = \frac{\rho g V H}{\eta \cdot 3600 s/h}
\end{equation}
where $E_{hour}$ is hourly pumping energy (Wh), $V$ is the volume pumped that hour (m³).

\subsubsection{Solar Resource Integration}

PV generation is modeled from irradiance data:

\begin{equation}
P_{PV} = G \cdot A \cdot \eta_{PV}
\end{equation}
where $P_{PV}$ is PV power output (W), $G$ is incident solar irradiance (W/m²), $A$ is PV panel area (m²), and $\eta_{PV}$ is the dimensionless PV conversion efficiency. System losses (inverter, temperature, wiring) follow standard PV modeling methodologies. At hourly time steps (8760 h/year), the model computes climate-adjusted demand, pumping energy, PV generation, and net load.

\subsubsection{Solar Geometry}

Solar geometry defines the sun's position relative to site latitude and day of year via the declination ($\delta$) and sunset hour angle ($\omega_{ss}$):
\begin{gather}
%\begin{equation}
\delta = 0.40928 \cdot \sin\left(\frac{2\pi(doy-80)}{365}\right) \\
%\end{equation}
%\begin{equation}
\omega_{ss} = \cos^{-1}(-\tan L \cdot \tan\delta)
%\end{equation}
\end{gather}
where $\delta$ is solar declination (rad), $\omega_{ss}$ is the sunset hour angle (rad), $doy$ is the day of year (1–365), and $L$ is site latitude (rad) \cite{pveducation, alamemon2022}.

\subsubsection{Hourly Irradiance Distribution (Markov Chain)}

Daily insolation is decomposed into hourly irradiance using Markov chain methods. The clearness index ($K_t$) is modeled as a random process with transition matrices from monthly averages:

\begin{equation}
K_t = \frac{H_t}{H_{et}}
\end{equation}
where $K_t$ is the dimensionless daily clearness index, $H_t$ is daily global surface insolation (kWh/m²), and $H_{et}$ is daily extraterrestrial insolation (kWh/m²). Transition matrices define the probability of moving between clearness states, ensuring realistic autocorrelation and variability \cite{aguiar1988}.

\subsubsection{Erbs Model Decomposition}

Global horizontal irradiance (GHI) is decomposed into diffuse ($G_d$) and direct beam ($G_b$) components:

%\begin{gather}
\begin{equation}
k_d = \begin{cases} 
1-0.09K_t, & K_t \le 0.22 \\
\begin{aligned}
&0.9511-0.1604K_t+4.388K_t^2 \\
&-16.638K_t^3+12.336K_t^4,
\end{aligned} & 0.22 < K_t \le 0.80 \\
0.165, & K_t > 0.80 
\end{cases} %\\
\end{equation}
\begin{equation}
G_d = k_d \cdot GHI, \quad G_b = GHI - G_d
\end{equation}
%\end{gather}
where $k_d$ is the dimensionless diffuse fraction, $GHI$ is global horizontal irradiance (W/m²), and $G_d$ and $G_b$ are the diffuse and direct beam horizontal components (W/m²) \cite{erbs1982}.

\subsubsection{Transposition Model}

Irradiance on tilted PV surfaces combines direct, diffuse, and reflected components:

\begin{equation}
G_{tilt} = G_b \cdot \frac{\cos\theta}{\sin h_s} + \left(\frac{1+\cos\beta}{2}\right) G_d + \left(\frac{1-\cos\beta}{2}\right) GHI \cdot \rho_g
\end{equation}
where $G_{tilt}$ is total tilted-plane irradiance (W/m²), $\theta$ is the angle of incidence (rad), $h_s$ is solar altitude (rad), $\beta$ is panel tilt (rad), and $\rho_g$ is the dimensionless ground albedo \cite{erbs1982}.

\begin{figure*}
\centering
\includegraphics[width=\textwidth]{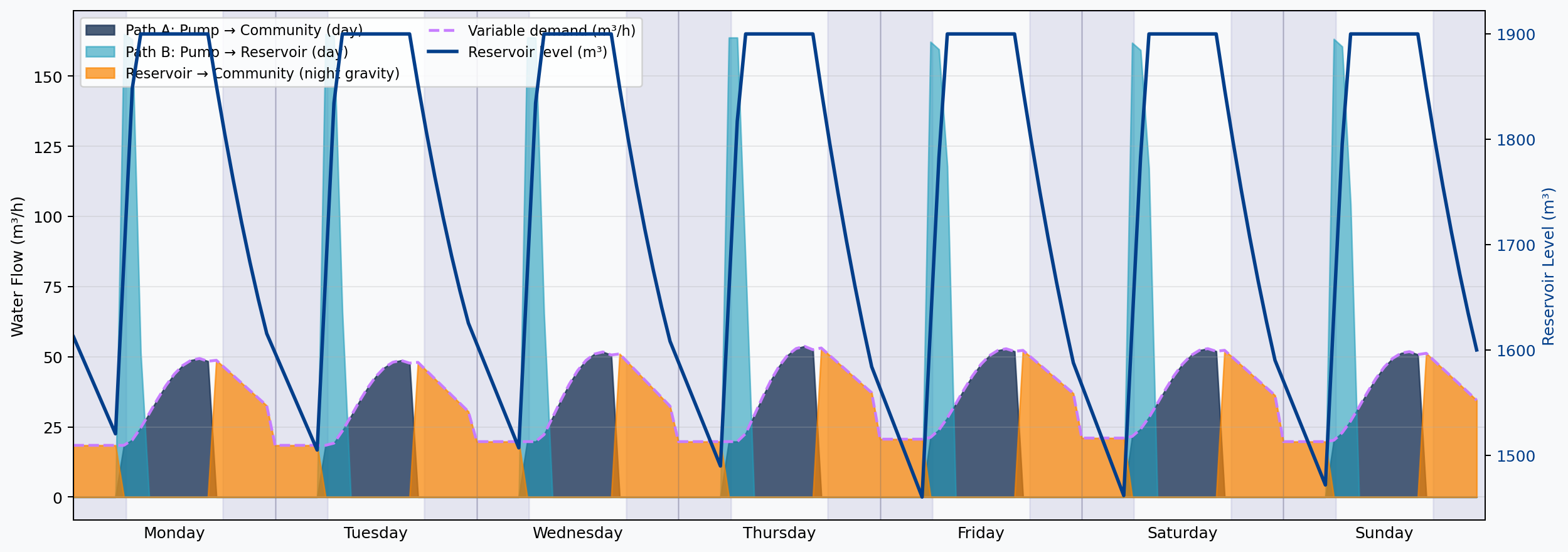}
\caption{Hourly water paths from the Python simulation for a sample week, showing pumped inflow, reservoir storage volume, and delivered demand.}
\label{fig:waterpaths}
\end{figure*}

\subsubsection{Hourly Temperature Model}

Hourly ambient temperature is simulated from daily max/min temperatures and solar radiation via the cumulative irradiation ratio ($k_x$):
\begin{gather}
%\begin{equation}
k_x(t) = \frac{\int_{sunrise}^{t} GHI(\tau) \, d\tau}{\int_{sunrise}^{sunset} GHI(\tau) \, d\tau} \\
%\end{equation}
%\begin{equation}
T_a(t) = \begin{cases}
T_{min} + slp_{before} \cdot k_x(t), & t \le t_{kxmax} \\
T_{max} - slp_{after} \cdot (k_{xmax} - k_x(t)), & t > t_{kxmax}
\end{cases} \\
%\end{equation}
%\begin{equation}
slp_{before} = \frac{T_{max} - T_{min}}{k_{xmax}}, \quad slp_{after} = 1.7 \cdot slp_{before}
%\end{equation}
\end{gather}
where $k_x(t)$ is the dimensionless cumulative irradiation ratio (0 at sunrise, 1 at sunset), $T_a(t)$ is ambient temperature (°C), $T_{min}$ and $T_{max}$ are daily min/max temperatures (°C), $k_{xmax}$ is $k_x$ at peak temperature, and $t_{kxmax}$ is the time of that peak, while $slp_{before}$ and $slp_{after}$ are heating/cooling slopes (°C per unit $k_x$). Nighttime cooling is modeled linearly between sunset and sunrise \cite{meteonorm2023}.

\subsection{Reservoir as Equivalent Battery Energy Storage System}

\subsubsection{The "Hydraulic Battery" Concept}

This study uses the term \emph{hydraulic battery} to mean the use of an existing elevated reservoir as a gravitational energy-storage medium functionally equivalent to an electrochemical battery for load shifting. Energy is "charged" by pumping water uphill when solar power is abundant and "discharged" by gravity-fed release when it is not. Conceptually, this is a form of pumped-hydro energy storage (PHES), but it differs in three ways. First, scale and purpose: conventional PHES couples two large reservoirs to a reversible turbine producing tens to hundreds of MW of electricity, whereas our hydraulic battery stores on the order of hundreds of kWh and "discharges" as \emph{water service} rather than re-electrification. Second, infrastructure reuse: the storage asset already exists in the water-supply system, so no dedicated reservoir pair or turbine hall must be financed. Third, conversion path: stored energy is consumed directly as delivered water—the pump is the only converter—avoiding the round-trip electricity-to-electricity losses of PHES. Comparable concepts appear globally, including micro-PHES retrofits of existing water infrastructure, "water-battery" demonstrations that schedule municipal pumping to absorb surplus renewables, and solar water-pumping with elevated tanks deployed across South Asia and Sub-Saharan Africa.

\subsubsection{Energy Storage Derivation and Parameter Estimation}

The reservoir is modeled as gravitational energy storage:
%\begin{gather}
\begin{equation}
E = \rho g h V %\\
\end{equation}
\begin{equation}
E_{kWh} = \frac{\rho g h V}{3.6 \times 10^6 \cdot \eta}
\end{equation}
%\end{gather}
where $E$ is stored potential energy (J), $E_{kWh}$ is electrical-equivalent energy (kWh), $h$ is effective storage head (m), $V$ is stored volume (m³), $\eta$ is combined pump/recovery efficiency (dimensionless), and $3.6 \times 10^6$ converts J to kWh. The battery-equivalent parameters used in Xendee were therefore not assumed but derived from physical reservoir specifications: effective head $h = 30$ m (elevation above the point of use), usable volume $V = 1900$ m³, and a resulting energy intensity of 0.163 kWh/m³ (obtained by evaluating $E_{kWh}/V$):

\begin{equation}
E_{res} = 1900 \times 0.163 \approx 311 \text{ kWh}
\end{equation}
where $E_{res}$ is total electrical-equivalent capacity (kWh). The charge/discharge efficiencies supplied to Xendee correspond to the pump efficiency (charging) and gravity-fed delivery efficiency (discharging), and the state-of-charge limits correspond to the tank's min/max operational water levels, aligning with established pumped-storage formulations \cite{harby2013}. In Xendee the reservoir is thus modeled as a BESS of 311 kWh capacity, charged by pump operation and discharged by gravity-fed supply, enabling load shifting so excess solar energy is stored as water for later use.

\subsection{Xendee Time-Series Optimization}

The optimization minimizes total system cost:

\begin{equation}
\min \sum_{t=1}^{T} \left( C_{grid}(t) + C_{O\&M}(t) \right)
\end{equation}
where $C_{grid}(t)$ is the cost of grid electricity purchased in hour $t$ (\$), $C_{O\&M}(t)$ is the operation and maintenance cost (\$), and $T$ is the number of timesteps. It is subject to power balance and storage dynamics:
%\begin{gather}
\begin{equation}
P_{PV}(t) + P_{grid}(t) = P_{load}(t) %\\
\end{equation}
\begin{equation}
SOC(t+1) = SOC(t) + \eta_c P_{charge}(t) - \frac{P_{discharge}(t)}{\eta_d}
\end{equation}
%\end{gather}
where $P_{PV}(t)$, $P_{grid}(t)$, $P_{load}(t)$ are PV, grid, and load power (kW); $SOC(t)$ is the reservoir state of charge (kWh-equivalent water level); $\eta_c$, $\eta_d$ are dimensionless charge/discharge efficiencies; and $P_{charge}(t)$, $P_{discharge}(t)$ are charging (pumping) and discharging (gravity-fed) power (kW). Grid outages are modeled by forcing $P_{grid}(t) = 0$, after which reliability is evaluated from stored water (reservoir SOC) and available PV.
\begin{figure*}
\centering
\includegraphics[width=\textwidth]{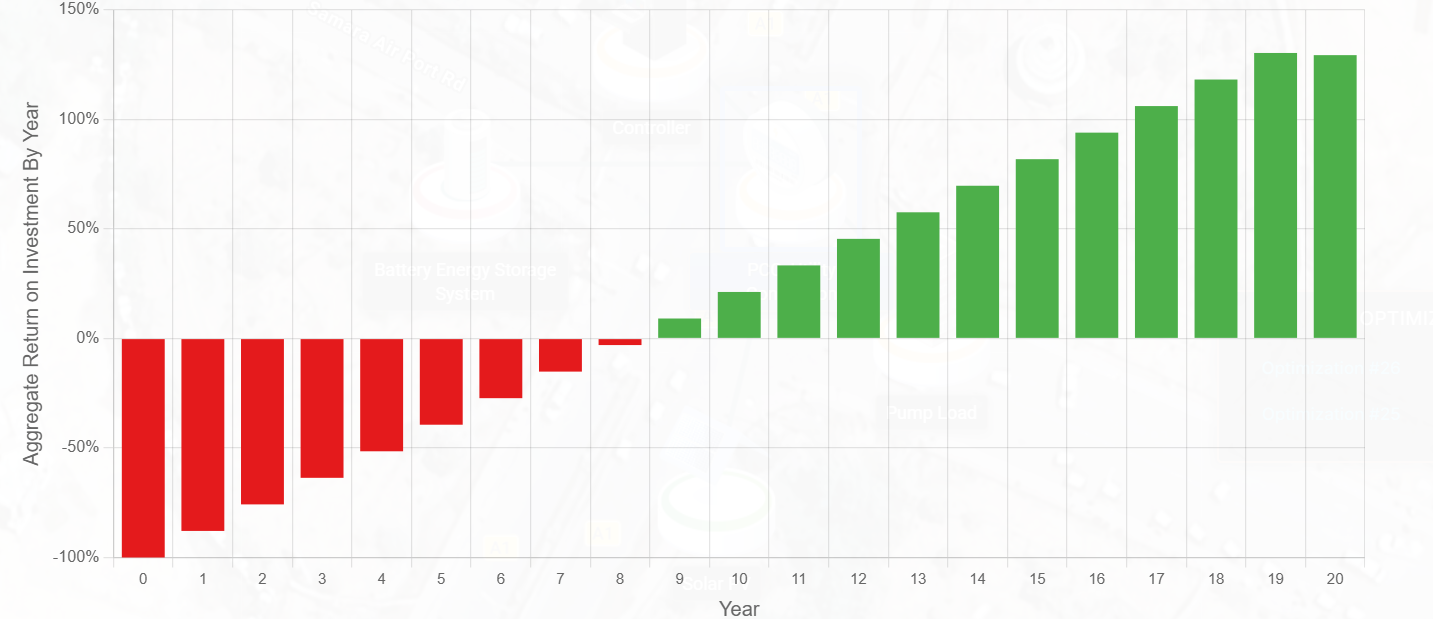}
\caption{Annual return on investment (\%) projected across the first twenty years of operations.}
\label{fig:cashflow}
\end{figure*}

\subsection{Economic Analysis}

Net present value (NPV), internal rate of return, and break-even are computed as:
%\begin{gather}
\begin{equation}
NPV = \sum_{t=0}^{N} \frac{R_t - C_t}{(1+r)^t} %\\
\end{equation}
\begin{equation}
0 = \sum_{t=0}^{N} \frac{R_t - C_t}{(1+IRR)^t} %\\
\end{equation}
\begin{equation}
\sum_{t=0}^{t_b} (R_t - C_t) = 0
\end{equation}
%\end{gather}
with $NPV$ in USD, $R_t$ and $C_t$ are revenue and cost in year $t$ (\$), $r$ is the discount rate (dimensionless), $N$ is the project lifetime (years), $IRR$ is the internal rate of return (the discount rate at which $NPV=0$), and $t_b$ is the break-even year (the first year cumulative net cash flow is non-negative). System adequacy during a blackout is ensured when:

\begin{equation}
E_{PV}(t) + E_{res}(t) \geq E_{load}(t)
\end{equation}
where $E_{PV}(t)$, $E_{res}(t)$, $E_{load}(t)$ are available PV energy, reservoir-stored energy, and required load energy (kWh) at time $t$. The reservoir enhances resilience by decoupling water supply from real-time energy availability.

\section{RESULTS}

\subsection{Technical Performance}

The Python simulation (covering 2,251 days) yielded a solar fraction of 55.1\% and delivered a mean of 944.4 m³/day, with the grid providing backup primarily during cloud gaps or nocturnal reservoir depletion. Fig.~\ref{fig:waterpaths} shows the hourly water paths over a representative week: pumped inflow tracks solar availability during daylight while the elevated reservoir buffers delivery so demand is met continuously, including overnight when no pumping occurs. The complementary filling and draining of the reservoir visible in the figure is the physical mechanism behind the load-shifting behavior.

\subsection{Economic Viability}

The total revised CAPEX is estimated at \$879,912 and annual OPEX at \$9,500. The Xendee model predicts an IRR of 10\%, a break-even point of 9 years, and an optimal water tariff of 12 Birr/m³ (\$0.076/m³), which ensures bankability while remaining competitive against trucked water that often exceeds 20 Birr/m³. The return on investment is plotted by year in Fig.~\ref{fig:cashflow}, with profit beginning in year 9. The "hydraulic battery" approach performs better in cost and longevity than electrochemical batteries, which suffer short lifespans (5–15 years) and high replacement costs.

\subsection{Reliability}

The model demonstrates that water storage can effectively replace batteries in outage scenarios: the load is fully met through a 24-hour utility grid outage on June 16, 2025 with no break in service, highlighting the reliability and cost effectiveness of the design.

\section{Discussion}

The results demonstrate that the proposed solar–grid hybrid water supply system meets its objectives of improving technical performance, economic feasibility, and resilience through an integrated modeling framework.

The 55.1\% solar fraction indicates that more than half of energy demand is met renewably, while the 944.4 m³/day average satisfies the demand profile with operational stability. Reliance on grid power during cloud cover and nighttime reservoir depletion highlights the value of combining storage with backup. The results confirm that reservoir-based storage effectively shifts consumption, allowing pumping to occur preferentially during peak solar hours—behavior shown in Fig.~\ref{fig:waterpaths} and consistent with the design objective of minimizing grid dependency while ensuring continuous supply.

Economically, the CAPEX of $\sim\$$879,912 with low annual OPEX ($\sim\$$9,500), a 10\% IRR, and break-even in year 9 indicate financial sustainability under realistic tariffs. Because the storage asset is an existing reservoir rather than a chemical battery, the project avoids both the upfront battery cost and the recurring replacement costs every 5–15 years—a primary driver of the favorable economics. On resilience, the system fully rides through the 24-hour outage without interruption; decoupling water delivery from instantaneous electricity supply means even a full-day grid loss does not interrupt water service, provided the reservoir is adequately charged—valuable in a region with a highly unreliable grid.

Methodologically, integrating Markov-chain stochastic simulation (Python) with deterministic time-series optimization (Xendee) offers an advantage over either method alone. The meteorological model represents solar, demand, and environmental uncertainty and yields statistically robust metrics over long horizons, while Xendee provides optimal operational strategies and high-resolution dispatch and economic trade-offs under cost and system constraints. Together they ensure the design is robust under uncertainty yet operated at minimum cost, producing a model closer to real-world operation.

Several limitations suggest future work. The analysis assumes a fixed reservoir geometry and a single site; extending to multiple sizes and elevations would clarify how the hydraulic-battery economics scale. The outage analysis considers a 24-hour event; multi-day scenarios and a full probabilistic reliability assessment (e.g., loss-of-load probability) would strengthen the resilience conclusions. Finally, optimizing feed-in or water-tariff agreements could further enhance attractiveness to private investors.

\section{CONCLUSION}

This study models a solar-grid hybrid water supply system for Semera, Ethiopia, using a dual-framework approach of a meteorological model and Xendee optimization. The results confirm that treating water reservoirs as energy-storage equivalents (approximated at 311 kWh) is technically viable and more affordable than traditional batteries, significantly reducing total CAPEX. With a consistent solar fraction of 55.1\% and a financially sustainable IRR of 10\%, the model demonstrates that hybrid systems can provide reliable, cost-effective water access in arid regions. By deriving storage parameters directly from reservoir physics and selecting an optimization platform suited to exogenous load profiles and storage dispatch, the work offers a transferable methodology for similar infrastructure-reuse projects. Future work should optimize feed-in tariff agreements and extend the reliability analysis to multi-day outage scenarios.

\end{document}